# Estimation of the in-situ elastic constants of wood pulp fibers in freely dried paper via AFM experiments

C. Czibula[1,2,6,#], A. Brandberg[2,3,#*], M. J. Cordill[4], A. Matković[1], O. Glushko[5], Ch. Czibula[6], A. Kulachenko[2,3], C. Teichert[1,2], U. Hirn[2,6]

[1]Institute of Physics, Montanuniversitaet Leoben, Franz Josef Str. 18, 8700 Leoben, Austria

[2]Christian Doppler Laboratory for Fiber Swelling and Paper Performance, Graz University of Technology, Inffeldgasse 23, 8010 Graz, Austria

[3]Solid Mechanics, Department of Engineering Mechanics, KTH Royal Institute of Technology, Teknikringen 8D, 114 28 Stockholm, Sweden

[4]Erich Schmid Institute for Materials Science, Austrian Academy of Sciences, Jahnstr. 12, 8700 Leoben, Austria

[5]Chair of Materials Physics, Department Materials Science, Montanuniversitaet Leoben, Jahnstr. 12, 8700 Leoben, Austria

[6]Institute of Bioproducts and Paper Technology, Graz University of Technology, Inffeldgasse 23, 8010 Graz, Austria

# equally contributing

* corresponding author

## Abstract

Atomic force microscopy-based nanoindentation (AFM-NI) enables characterization of the basic mechanical properties of wood pulp fibers in conditions representative of the state inside a paper sheet. Determination of the mechanical properties under different loads is critical for the success of increasingly advanced computational models to understand, predict and improve the behavior of paper and paperboard. Here, AFM-NI was used to indent fibers transverse to and along the longitudinal axis of the fiber. Indentation moduli and hardness were obtained for relative humidity from 25 % to 75 %. The hardness and the indentation modulus exhibit moisture dependency, decreasing by 75 % and 50 %, respectively, over the range tested. The determined indentation moduli were combined with previous work to estimate the longitudinal and transverse elastic modulus of the fiber wall. Due to the relatively low indentation moduli, the elastic constants are also low compared to values obtained via single fiber testing.

## Keywords

*AFM-based nanoindentation, wood pulp fiber, S2 layer, anisotropy*

# Introduction

Paper is an essential part of daily life and is becoming more important with the development of a bio-based sustainable economy. When normalized with respect to density, the mechanical properties of paper and paperboard originate mainly from fiber and fiber-fiber-bond quality [1]. These properties exhibit large variance and are time-consuming to test due to the challenges of isolating, manipulating, and restraining individual fibers [2][3][4][5] and fiber joints [6][7]. The single pulp fiber is difficult to test in compression or using the elastica loop method [8]. A recent review of the challenges in computational modeling of paper and paperboard notes the difficulty of obtaining mechanical characterization data on the fiber level as one of the problems in the field [9]. Nanoindentation (NI) can complement traditional characterization experiments as the indentation is very local and loads the material mainly in compression. Loading the sample in compression reveals the effect of some mechanisms that are hidden in tension such as fiber porosity [10] and nanofibril buckling (as suggested in [11]). These mechanisms potentially affect the response of the fiber and, therefore, of the paper sheet when loaded in compression or bending.

The pulp fiber is an industrially processed wood fiber, which is freed from the wood by mechanical, thermal, and/or chemical treatment. Wood fibers have a complex hierarchical structure and anisotropic material's properties [12]. The cell wall of individual wood fibers is composed of cellulose microfibrils which are surrounded by a matrix of an amorphous material (mainly hemicellulose and lignin) and consists of different layers (primary (P) layer, secondary (S1, S2, and S3) layers) that differ in thickness, chemical composition, and microfibril alignment. Typically, the fiber is hollow with an open lumen (L). The orientation of the microfibrils is characterized by the microfibril angle (MFA) θ, relating the orientation of the fibrils to the longitudinal fiber axis. The S2 layer is the thickest and dominates the mechanical properties of the fiber, especially in the axial direction. In this layer, the cellulose microfibrils are aligned close to the axis of the fiber and the MFA is relatively low. During the pulping process, the wood fiber undergoes several structural changes, as illustrated in Figure 2. Often, the P layer is removed due to its high lignin content and random fibril alignment, and the lumen collapses during drying and sheet formation [13–15]. The fiber may also become fibrillated and partially damaged by mechanical action, and the lignin may be chemically removed. Hence, the pulp fiber at the onset of sheet drying is a complex structure with large variations from fiber to fiber.

Paper and paperboard exhibit significant variation in mechanical performance depending on the restraint imposed during drying [16]. Wuu [17] showed that the elastic modulus of fibers surgically extracted from handsheets dried freely was less than half of the same type of fibers extracted from the headbox. Kappil and co-authors reported that inside sheets that have been dried under restraint in one direction and freely in the other, there is a clear difference in mechanical properties between fibers oriented mainly in the machine direction and those oriented in the cross direction [7]. Both Wuu and Kappil et al. determined the fiber properties using a combination of optical and micro-mechanical testing. However, surgically removing the fibers from the sheet is laborious and error prone. NI is an attractive alternative to micromechanical testing as the sheet can be made and dried under industrially relevant conditions and then cut by a microtome to expose the in-plane fiber cross-sections. NI experiments along the longitudinal axis of the fiber have been reported for both wood fibers [18–23] and pulp fibers [24,25]. The influence of relative humidity on the indentation modulus and hardness of wood fibers has also been reported [26,27].

An alternative to performing conventional NI is to employ an atomic force microscope (AFM), effectively repurposing the tip at the end of the cantilever beam, which is normally used to map the surface topology, to act as an indenter. This approach is attractive as morphological information and mechanical characterization can be extracted with the same device. For wood pulp fibers, AFM-based methods enable access to the transverse direction by overcoming the high surface roughness of the fiber surface and allow the implementation of more complex experimental approaches to study mechanical properties [28,29] as well as fiber bonding mechanisms [30,31]. Inspired by NI studies, AFM-based nanoindentation (AFM-NI) has been adapted to study the influence of relative humidity on cellulosic fibers in the transverse direction and films [32–34] very locally. Similar methods have also been employed to study cortical bone, which shares several traits with wood such as orthotropic and moisture dependent stiffness [35].

It is important to extend the study of the mechanical behavior of single pulp fibers to different relative humidity levels since the elastic properties of paper sheets decrease at higher moisture content [12,36][37]. This decrease stems from both inter- and intra-fiber mechanisms, most importantly the fiber-fiber bonds which tend to dissolve at higher moisture content and the amorphous regions of cellulose which are susceptible to softening [38].

In this work, we investigated the effect of relative humidity on the indentation modulus in the axial direction of the fiber and compare it with the indentation modulus in the transverse direction. We combined the results of both directions to determine the elastic stiffness tensor components under the assumption that the fiber can be described as a transversely orthotropic viscoelastic material. The results were compared with previous NI studies and results obtained using traditional micro-scale single fiber testing. Such a comparison reveals the influence of mechanisms exclusive to compression and tension.

## Materials & Methods

### *Microtome & sample preparation*

Microtome cuts of an industrial, unbleached, and unrefined pulp (Mondi Frantschach, kappa number – indicating the residual lignin content of the pulp – κ = 42) were prepared. First, the paper was embedded in a hydrogel-like material, glycol methacrylate (GMA) [39,40], and the paper was then cut by a diamond knife to a slice thickness of about 7 µm. For AFM measurements, these 7 µm thick microtome cut slices of paper were attached to a steel sample holder with nail polish – analogous to [32][41]. For comparison with the transverse direction, six individual fibers of the same pulp (Mondi Frantschach, κ = 42) were used from a previous study, thoroughly described in [29].

### *Raman spectroscopy*

Raman analysis was performed in a confocal configuration using a Horiba Labram HR Evolution spectrometer. Raman spectra were recorded with a 532 nm laser (50 mW power on the sample surface), a 600 l/mm grating, a 100x magnification lens, 6 s acquisition times, and 10 accumulations. In the presented spectra, cosmic rays were removed, the background was deduced, and intensities were normalized by the Labspec 6 software. Using higher laser power or significantly longer acquisition times was found to introduce damage to the sample. To ensure that there was no laser-induced damage, multiple Raman spectra were measured from the same spots.

### Confocal laser scanning microscopy

A confocal laser scanning microscope (CLSM) Olympus LEXT 4100 OLS was used for optical 2D and 3D imaging of the fiber cross-section samples. A laser wavelength of 405 nm was used, and the lateral resolution was 120 nm. The z-resolution is theoretically as small as 10 nm thanks to a multi-focusing algorithm.

### Nanoindentation (NI)

Nanoindentation was performed on a Bruker-Hysitron TS 77 Select using a well-calibrated cube corner tip. Appropriate indentation sites were identified by scanning the tip over the prepared sample surface. Once appropriate areas were found, the indentation mapping method was used to make 50 indents with a maximum load of 100 µN. The loading profile was 1 s load – 10 s hold – 1 s unload. The unloading slope and the Oliver and Pharr method [42] were used to evaluate the hardness and reduced elastic modulus as described in the next section. As a comparison, indents using the same method and load were made into the embedding material GMA.

### AFM

All AFM based nanoindentation (AFM-NI) measurements were performed with an Asylum Research MFP-3D AFM. The instrument is equipped with a closed-loop planar x–y-scanner with a scanning range of 85 x 85 µm² and a z-range of about 15 µm. For topography imaging, AC160TS-R3 silicon probes (Olympus, Japan) with an aluminum reflex coating on the backside of the cantilever and a nominal tip radius of 10 nm were used.

Two different probe geometries were chosen for AFM-NI. First, pyramid-shaped ND-DYIRS probes (Advanced Diamond Technologies, USA) were used. These probes are full-diamond probes with aluminum reflecting backside coating and have a regular, four-sided pyramid as a tip. The front, back, and side angles are all 45°. The cantilever's spring constant was determined to be (82.1 ± 17.7) N/m using the thermal sweep method [43]. The thermal Q factor is 623 ± 54 and the resonance frequency is (378 ± 14) kHz (values are given as mean ± standard deviation calculated from four independent measurements). These probes were also used in [34] and were thoroughly discussed there.

Second, hemispherically shaped LRCH250 silicon probes (Team Nanotec, Germany) were used. The spring constant of the cantilever was (290.2 ± 51.3) N/m and was calibrated using the thermal sweep method (values are given as mean ± standard deviation calculated from four independent measurements). The thermal Q factor is 778 ± 224, and the resonance frequency is (575 ± 2) kHz. For the pyramidal and the hemispherical probes, the tip radii were found to be 100 nm and 300 nm, respectively. For the characterization of the tip, a calibration grid (NT-MDT, Russia) with sharp spikes was scanned and by taking advantage of the dilation principle [44], it was possible to image the tip itself. The 3D-AFM topography images of the indenter shapes are presented in Figure 1c for the hemispherical and pyramidal probe. Using these images and fitting the data with paraboloids, it was possible to obtain the tip radius. Here, it should be noted that the pyramidal probe used in this work were rather blunt and therefore, the cap of the pyramid had a hemispherical form which allowed the determination of the tip radius.

To investigate single fibers in an environment with controlled relative humidity, the AFM was equipped with a closed fluid cell (Asylum Research, USA) which can be flushed by nitrogen in a controlled way.

A setup for humidity control in the AFM was adapted from [45], and a more detailed description can be found in [32,34].

During the AFM-NI experiments, the surface to be tested was first scanned in tapping mode with a high-resolution probe (tip radius about 10 nm). For the longitudinal fiber direction, the surface of the fiber cross-section is rather smooth (about 15 nm root mean square (RMS) roughness). The transverse fiber direction exhibits a higher roughness (RMS roughness of about 150 nm), and it was more challenging to find suitable, flat regions on the surface. Next, force-indentation depth curves were recorded in contact mode at the desired position. For all measurements, the load schedule consists of one hold time of 10 s at maximum load and a second hold time of 30 s at the end of the unloading segment, as presented in Figure 1a. Loading and unloading were done at a constant load rate. The load level and load rate for the pyramidal and hemispherical probe were $F_{max}$ = 10 μN and $\dot{F}$ = 10 μN/s, and $F_{max}$ = 20 μN and $\dot{F}$ = 20 μN/s, respectively. During the hold times, the force was actively held at the target value.

The values for indentation modulus $M$ and the hardness $H$ are evaluated in this work according to the technique pioneered by Oliver and Pharr, where the instantaneous unloading slope $S_u = (dF/dz)|_{z_{max}}$ is used to estimate the elastic response [42]. The approach of Feng and Ngan is adopted to compensate for creep during the load ramp and the hold part of the experiment [46]. Instead of fitting $F(z)$, we fit $z(F)$, effectively determining the compliance $(dz/dF)|_{F_{max}}$. The first 75 % of the unloading curve were used to fit the constants $A_1, A_2, A_3, A_4$ in Equation (1).

$$z(F) = A_1 + A_2 F^{0.5} + A_3 F^{A_4} \tag{1}$$

The slope of unloading $S$ in Equation (4) is related to the observed slope of the unloading $S_u$ by Equation (2),

$$\frac{1}{S} = \frac{1}{S_u} + \frac{\dot{h}_h}{|\dot{F}|} \tag{2}$$

where $\dot{h}_h$ is the indenter displacement recorded at the end of the load hold time, and $\dot{F}$ is the unload rate at the start of the unloading. This method, mechanically equivalent to a linear spring-dashpot series system, has proven to be useful when the material exhibits a time-dependent response which distorts the measured slope during unloading unless it is corrected. The contact depth $z_c$ is calculated using Equation (3) in the way proposed by Oliver and Pharr [42] similarly corrected for the time dependence of the material according to the method in [47]. The parameter $\varepsilon$ was chosen 0.75 for both indenters. This approach of accounting for viscous effects is valid under the condition that the contact area is monotonically increasing up to the point of unloading [48].

$$z_c = z_{max} - \varepsilon \frac{F_{max}}{S} \tag{3}$$

The indentation modulus $M$ can then be found via Equation (4),

$$M = \frac{\sqrt{\pi}}{2\beta} \frac{S}{\sqrt{A(z_c)}} \tag{4}$$

where $S$ is the corrected slope of the unloading curve, which is fitted as illustrated in Figure 1b. $A(z_c)$ is the indenter's projected area at the contact indentation depth $z_c$. The area function $A(z)$ of an indenter describes the projected area at a certain distance $z$ from the tip apex [34]. The factor $\beta$ was set to 1.05 for both indenters [49][50]. The method suggested by Feng and Ngan was compared with the Oliver and Pharr model and yielded similar results. The method suggested by Feng and Ngan did however fit the unloading curve more closely at the start of unloading, as expected.

The hardness $H$ is calculated using Equation (5),

$$H = \frac{F_{max}}{A(z_c)} \tag{5}$$

where $F_{max}$ is the maximum applied force.

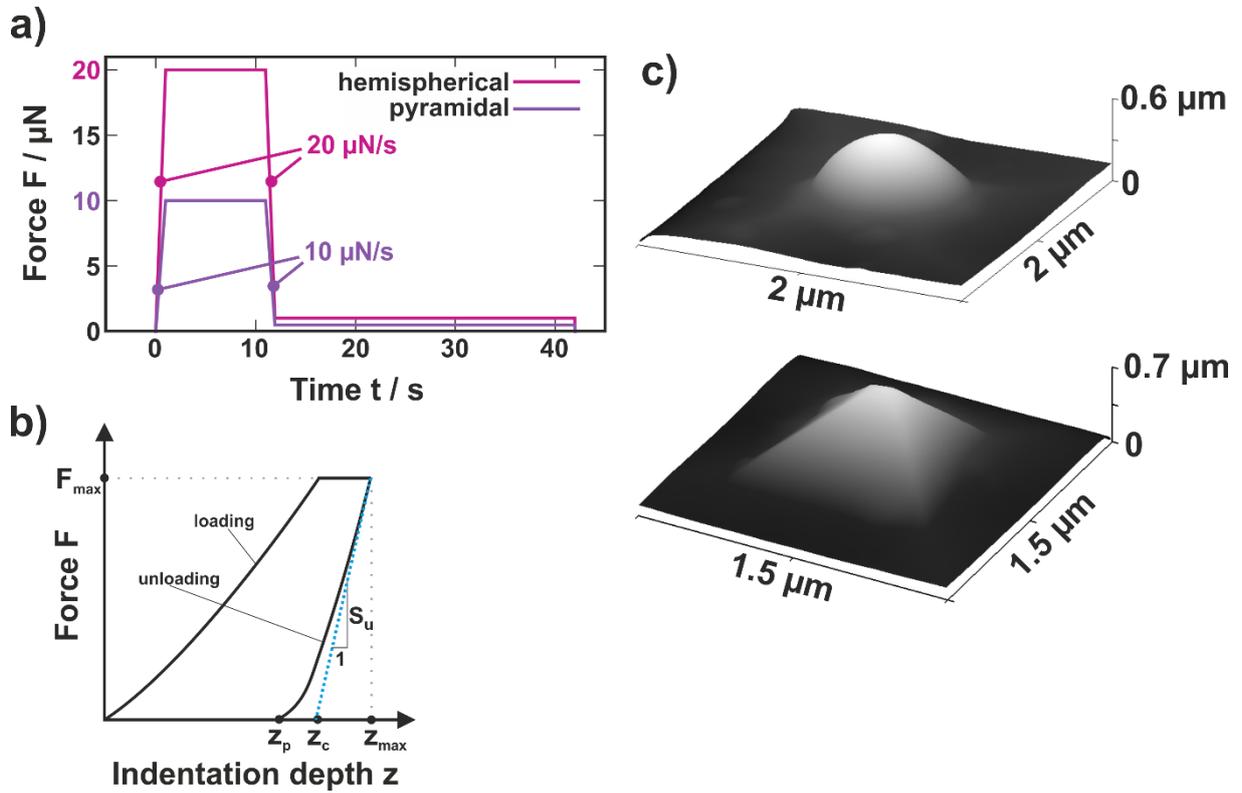

Figure 1: (a) Load schedule of the AFM-NI experiments for the hemispherical and pyramidal probe. (b) Force-Indentation depth plot illustrating the loading and unloading curve of the experiments. By fitting the slope S of the unloading curve, the indentation modulus M was calculated (Equation (4)). At maximum force F*max*, the hardness H of the surface was determined (Equation (5)). In (c), representative images of the hemispherical and pyramidal probe scanned with a calibration grid are illustrated.

*Relating indentation modulus and stiffness tensor components*

The indentation modulus $M$ obtained from experiments can be used to calculate the elastic moduli of the material. In isotropic materials, the relation between indentation modulus (sometimes referred to as the reduced modulus for isotropic materials) and elastic moduli $E, E_i$ is given by Equation (6),

$$\frac{1}{M} = \frac{1-\nu^2}{E} + \frac{1-\nu_i^2}{E_i} \tag{6}$$

where the subindex $i$ refers to the indenter material properties and $\nu, \nu_i$ are the Poisson's ratios of the material under investigation and the indenter material, respectively [42]. Such Hertz contact analysis assumes frictionless interaction, smooth surfaces, and small deformations relative to the radius of the indenter and the size of the bodies in contact.

Since wood fibers are anisotropic, the indentation modulus cannot be directly interpreted as an elastic stiffness component (e.g., the longitudinal Young's Modulus $E_L$) [21]. Due to the complex stress-state underneath the indenter, the measured modulus represents a mix of stiffness components. Relating the indentation modulus to the anisotropic elastic stiffness components is essential to compare NI and micromechanical test results [51,52], as will be discussed below.

The fiber wall is usually modeled as a transversely isotropic material with the axis of symmetry coinciding with the MFA orientation. Such materials are described by five independent material constants. In this case, the indentation modulus is a function of all five parameters, as well as the orientation of the principal material axes relative to the indentation directions [52]. Jäger and co-authors adapted the framework proposed by Vlassak et al. [53] to the problem of relating the indentation modulus of wood fibers to the orthotropic elastic components [51,52]. The method relies on parallel identification of the MFA to find the offset angle between indentation direction and material base and indentation measurements done at several angles relative to the longitudinal axis of the fiber [52]. The method was validated by performing the proposed indentations and identifying the unknown material parameters $E_L$, $E_T$, $G_{TL}$ [51]. The Poisson's ratios were set to fixed values and the remaining three parameters were determined by inverse modeling.

*Assumptions regarding mechanical behavior*

The theory presented by Vlassak et al. has been successfully used and validated on wood fibers and the results also compare reasonably well with macroscopic measurements using ultrasound [51,54,55]. Although derived with the Oliver-Pharr method in mind, the method does not consider plasticity nor viscoelasticity in the response. Instead, we compensate the influence of viscoelasticity during the determination of the indentation modulus (Equations (2) - (4)). The influence of plasticity on the measured response is unknown, even if the unloading (where the slope is estimated) is fully elastic, due to the potential for pile-up or sink-in phenomena around the indenter. In the ESI, section S6, the effect of viscoelasticity is investigated with a FEM model, and the assumptions are shown to be reasonable.

A second concern is how to account for the outer S1 layer. The S1 layer has fibrils oriented in the plane, but more randomly than the S2 layer. We assume that indentations into the S1 layer can be considered as indenting along one of the weak axes in the S2 layer, as in both cases there are no fibrils oriented in the direction of the indentation, as illustrated in Figure 2. In the absence of a more detailed description of the S1 layer's fibril orientations and keeping in mind that the low thickness may justify the use of a film-on-substrate type of analysis, we do not make any attempts to model this aspect of the fiber.

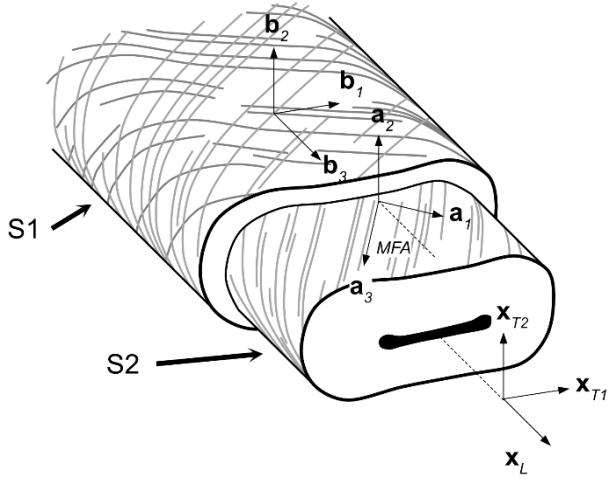

*Figure 2: Schematic illustration of a wood pulp fiber. Most of the fiber volume is present in the S2 layer, while the S1 layer is forming an outer shell. During the papermaking process, the S1 layer may sustain significant damage, partially exposing the underlying S2. The fiber is characterized by the transversely orthotropic base $\{x_{T1}, x_{T2}, x_L\}$. Relative this base, the microfibrils are aligned at some angle, which is typically between 0 and 40 degrees, illustrated here via the angle between the axis $x_L$ and $a_3$. The S2 layer is transversely orthotropic with the axis of symmetry (the strong direction) oriented along $a_3$. The S1 has a less well-defined fibril orientation, which is here represented by two different orientations of the fibrils in the plane of the fiber surface. The resulting material is similarly transversely isotropic with $b_2$ as the axis of symmetry (here, the weak direction).*

### *Identifying the longitudinal and transverse elastic modulus*

In this work, we employ a simple method to determine the material parameters from AFM-NI experiments performed only in two directions: longitudinally and transverse to the fiber direction. Implicitly, it is assumed that the MFA is 0 and that the longitudinal axis of the fiber coincides with the axis of symmetry for the transversely isotropic material. The neglect of the MFA allows the indentation modulus to be expressed as an explicit function of the stiffness tensor. Under such conditions, the approach of Vlassak et al. [53] can be solved using the method presented by Delafargue and Ulm [56]. The indentation modulus in the longitudinal direction of the fiber is given by the components of the material stiffness constants $C_{ijkm}$ as

$$M_L = 2\sqrt{\frac{C_{1111}C_{3333} - C_{1133}^2}{C_{1111}}} \left( \frac{1}{C_{2323}} + \frac{2}{\sqrt{C_{1111}C_{3333}} + C_{1133}} \right)^{-1} \tag{7}$$

while the indentation modulus in the transverse direction is given by Equation (8).

$$M_T = \frac{1}{2E(e)H_2^{3/4}H_3^{1/4}} \tag{8}$$

Here, $H_2$, $H_3$ and $e$ are given by Equations (9)-(11) and $E(e)$ is the complete elliptic integral of the second kind. This method has been successfully applied to identify the elastic moduli of bone [57]. Although [56] addresses only sharp indenters, a comparison with the theory of Vlassak et al. [53] indicates that the error is small when using the formula for indentations performed with spherical indenters. This comparison is shown in the ESI, section S5.

$$H_2 = \frac{1}{2\pi} \sqrt{\frac{C_{3333}}{C_{3333}C_{1111} - C_{1133}^2} \left( \frac{1}{C_{2323}} + \frac{2}{\sqrt{C_{3333}C_{1111}} + C_{1133}} \right)} \tag{9}$$

$$H_3 = \frac{1}{\pi} \cdot \frac{C_{1111}}{C_{1111}^2 - C_{1122}^2} \tag{10}$$

$$e = \begin{cases} \sqrt{1 - \dfrac{H_2}{H_3}} & \text{if} \quad H_2 < H_3 \\ \sqrt{1 - \dfrac{H_3}{H_2}} & \text{otherwise} \end{cases} \tag{11}$$

Equations (7) and (8) map the path between the elastic stiffness tensor $C_{ijkm}$ and the observed indentation modulus. To obtain the elastic stiffness components, we employed an inverse modeling scheme where a stiffness tensor was initially guessed and used to calculate the resulting indentation modulus in a certain direction. Through error minimization, it is then possible to determine the $C_{ijkm}$ which minimizes the error between observed moduli and calculated moduli. The minimization procedure (Equation (12)) was performed in MATLAB using the BFGS Quasi-Newton optimization scheme.

$$\min R = \frac{1}{n} \sum_{i=1}^{n} (M_i^{\text{model}} - M_i^{\text{exp}})^2 \tag{12}$$

The inputs to $M_i^{\text{model}}$ are described in the Results section of this work after the experimental results have been presented.

### *Relation to the sheet modulus*

In addition to the handsheets that were microtomed and used to perform AFM-NI, additional sheets were produced and used to measure the elastic modulus of the sheet. Six in-plane isotropic handsheets with a grammage of 80.4 ± 1.4 gm$^{-2}$ were produced with a Rapid-Köthen machine (DIN EN ISO 5269-2:2004), cut into four test specimens per sheet, and tested in tension with an L&W tensile tester (DIN-EN ISO 1924-2, 20 mm/min testing speed, 100 mm span length). The longitudinal modulus of the fiber is related to the modulus of the sheet. Several analytical formulas exist [58], the theory of Cox being the first presented [59]. Making several assumptions about the mechanical interactions inside the sheet, an upper bound on the elastic modulus of the sheet $E^{(sheet)}$ is given by Equation (13) where $E_L^{(fiber)}$ is the longitudinal modulus of the fiber and $\rho$ is the density of the sheet and the fiber, respectively.

$$E^{(sheet)} = \frac{1}{3} E_L^{(fiber)} \frac{\rho^{(sheet)}}{\rho^{(fiber)}} \tag{13}$$

## Pre- and postprocessing

The code used to produce the results in this work is available under an MIT license [60]. Tensor manipulations were done using the MMTensor toolbox originally developed for [61][62].The postprocessing and line art were prepared using MATLAB and the visualization toolbox GRAMM [63]. The CLSM and AFM images were analyzed by the Gwyddion software [64].

# Results

## Raman spectroscopy

A Raman spectra analysis has been performed to investigate whether GMA is penetrating the fiber cell wall. A comparison between the fiber and GMA Raman spectra is presented in Figure 3a. In the spectrum of the fiber, mainly characteristic cellulose modes (in the ranges of 300-500 $cm^{-1}$ and 1100-1400 $cm^{-1}$), a lignin characteristic mode at about 1600 $cm^{-1}$, and a $CH_2$-wagging mode (within 1440-1480 $cm^{-1}$ range) were observed [65]. In the case of GMA, the $CH_2$-wagging mode was also observed, and two additional modes at 603.4 $cm^{-1}$ and at 1725 $cm^{-1}$ which were characteristic only for the GMA [66]. Since these two modes were not observed in any of the spectra measured on the fibers, the results show that GMA does not penetrate the fiber cell wall. Therefore, it is assumed that the embedding of the fibers in GMA to produce microtome slices has a negligible influence on the mechanical properties of the fibers during the AFM-NI experiments.

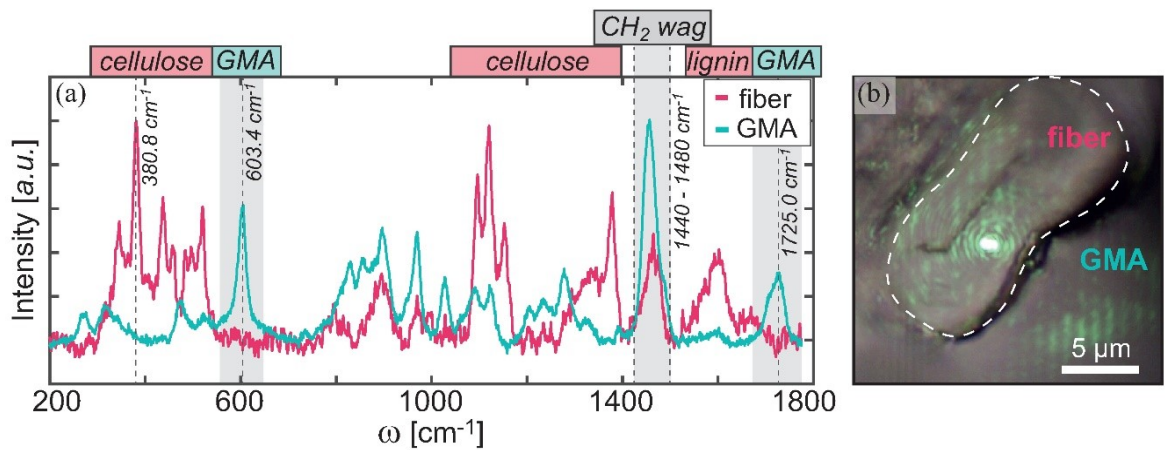

*Figure 3: (a) Raman spectra of a fiber and surrounding GMA. In the top, spectral ranges for the modes of cellulose, lignin, $CH_2$ wagging, and GMA are marked. Two characteristic modes of GMA (at 603.4 $cm^{-1}$ and 1725 $cm^{-1}$) that were not observed in the fibers are highlighted. $CH_2$ wagging mode (1440–1480 $cm^{-1}$ range) and a characteristic cellulose mode at 380.8 $cm^{-1}$ are also highlighted. (b) 20x20 $\mu m^2$ optical microscopy image of the fiber. The dashed white line marks the fiber perimeter for clarity. The laser spot on the sample is visible and marks the spot from which the Raman spectrum (a) of the fiber was measured. For the reported GMA spectrum, the laser spot was positioned in the bottom-right corner of (b).*

*Morphological Analysis*

Using confocal laser scanning microscopy (CLSM), the surfaces of the microtome cuts were investigated to obtain large overview images. In Figure 4, a CLSM laser intensity image with the corresponding height map is presented. Most clearly identifiable fibers have been cut perpendicular to the longitudinal fiber direction. Fibers that have been cut parallel to the longitudinal direction are not easily distinguishable. In Figure 4a, most of the visible fiber cross-sections exhibit a fully collapsed lumen, which is indicated by white arrows. Fibers that are suitable for mechanical characterization measurements are are marked by white dashed contours. A closer look at the individual fiber cross-sections in Figure 4a reveals the potential influence of the cutting procedure on the surface of the samples. Most cross-sections of fibers have a surface that contains cracks or fractures. These cracks appear in some cases perpendicular to the cutting direction. The cracks may be a result of prior treatment of the fibers. The height image in Figure 4b indicates that the embedding material GMA is usually located higher than the fiber cross-sections.

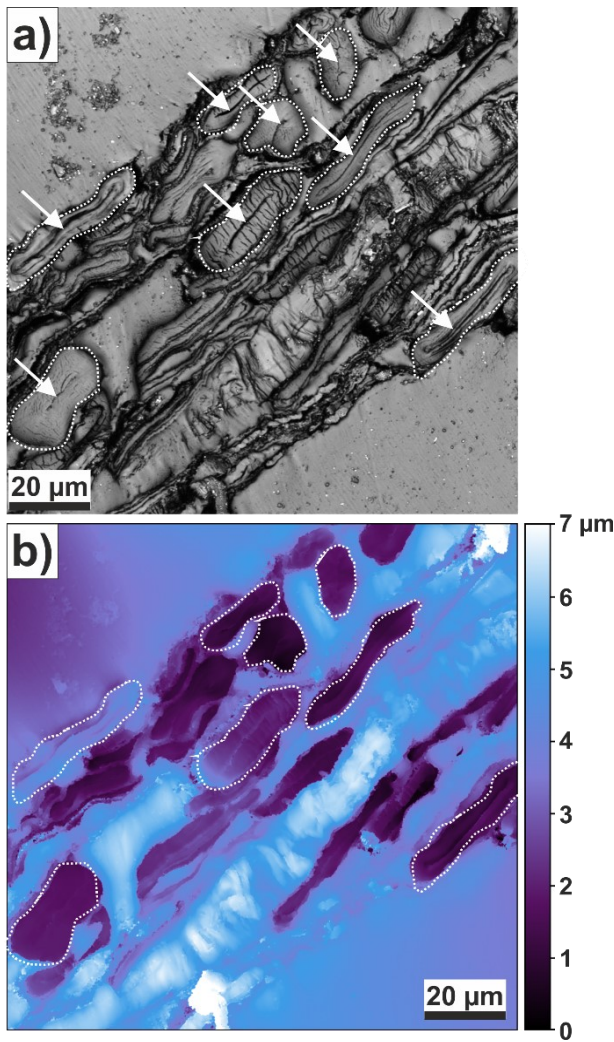

*Figure 4: Confocal laser scanning (CLSM) images. (a) CLSM laser intensity image of the surface of a microtome-cut slice of paper. (b) Corresponding height image (Z-scale: 7 µm). The white dashed contours mark fiber cross-sections which are considered isolated enough for AFM-NI measurements. Each white arrow in (a) indicates a collapsed lumen of a fiber.*

The observed surface features were also investigated with AFM. Topography images are presented in Figure 5. In Figure 5a, the surface of a fiber cross-section with fractures can be seen with zoom-ins in Figure 5b and c. The measurement of cross-sections – as presented in Figure 5e – indicates that the cracks have a width of about 100 nm and a depth of about 50 nm. In the AFM scans, the orientation of the cracks seems rather arbitrary.

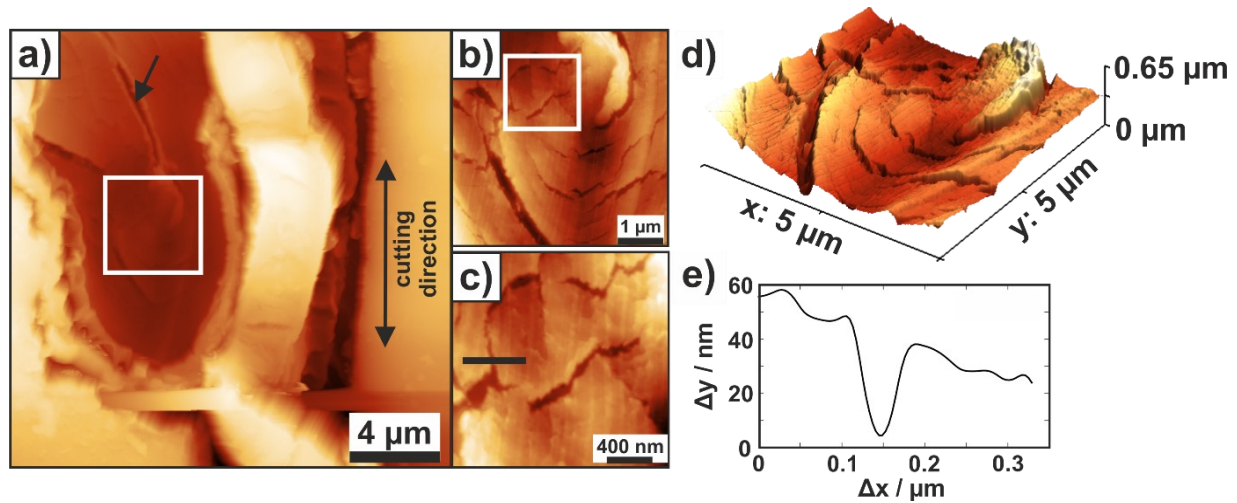

*Figure 5: AFM topography images. (a) 20 x 20 µm² topography image of a fiber cross-section cut perpendicular to the longitudinal fiber axis (z-scale: 4 µm). The black arrow points at the collapsed lumen whereas the white squares indicate the zoom-in regions. Zoom-ins: (b) 5 x 5 µm² topography image (z-scale: 650 nm) and (c) 1 x 1 µm² topography image (z-scale: 300 nm). The black line indicates a cross-section to determine the width and depth of the crack which is illustrated in (e). (d) 3D representation of the topography image in (b). (e) Cross-section of a crack as indicated by the black line in (c).*

### *AFM-NI results*

To mechanically characterize the S2 layer in the longitudinal direction, AFM-NI measurements were performed with the pyramidal and hemispherical probe at 25 %, 45 %, 60 %, and 75 % RH following the procedures described in [32,34]. A total of 777 successful indents were performed, 373 indents using the pyramidal indenter, and the remaining 404 using the hemispherical indenter. Box plots of the results for the indentation modulus and hardness are presented in Figure 6 and Figure 7, respectively, for both indenter shapes. The mean values along with standard deviations and confidence intervals are provided in Table 1, although – as can be seen in Figure 6 – the indentation modulus is not normally distributed. There is quite some scatter in the results, with the observed distributions exhibiting positive skewness, possibly due to the small volume tested during each indentation. The median values follow a clear trend from higher to lower indentation modulus for both indenters. The measurements generated several outliers, which are indicated by dots in Figure 6 and Figure 7. Not displayed are outliers with an indentation modulus larger than 20 GPa but less than 50 GPa, which were few but occurred for all relative humidity levels. Indentation moduli above 50 GPa were directly discarded as failures of the postprocessing algorithm, as were measurements that failed to produce a positive unloading slope even after correcting for the creep at the holding load. These failures may occur if the thermal drift rate of the system is not constant, or due to the noise present in the measured signal. The results for the indentation modulus in Figure 6 indicate that the fiber wall may lose up to half of its indentation modulus locally as the relative humidity is increased from 25 % to 75 %. The indentation exhibits a trend similar to that observed by Wanju et al. [27] although the absolute magnitudes measured are lower. Both indenter shapes show a similar response. In the electronic supplementary information (ESI), representative experimental curves for both indenter shapes are illustrated in Figure A1.

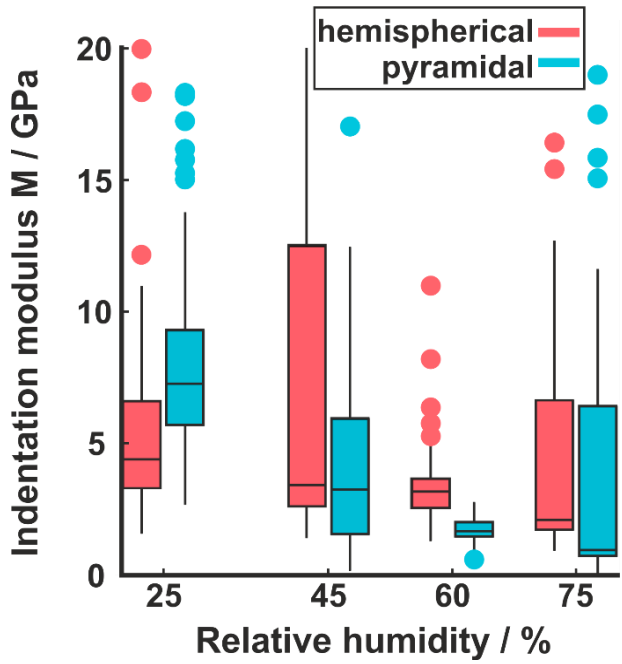

*Figure 6: Results for the indentation moduli of the fibers obtained from the AFM-NI experiments with both indenter shapes in the longitudinal direction at 25 %, 45 %, 60 %, and 75 % relative humidity. The horizontal bar indicates the median of the group, the box is the interquartile range, the whiskers cover the range of data not considered outliers and the dots are outliers. The criterion to classify an observation as an outlier is that it is greater than $q_3 + 1.5 \cdot (q_3 - q_1)$ or less than $q_1 - 1.5 \cdot (q_3 - q_1)$, where $q_1, q_3$ are the 25$^{th}$ and 75$^{th}$ percentiles of the sample data, respectively.*

*Table 1: Statistics of the evaluation of the AFM-NI experiments at different relative humidity for the pyramidal and hemispherical indenter. N – number of samples in the set, M – Indentation modulus, H – Hardness. Values are given as the estimated mean ± confidence interval at confidence level $\alpha = 0.05$ with the sample standard deviation given in brackets.*

|  |  | $N$ | $M$ [GPa] | $H$ [GPa] |
|---|---|---|---|---|
| *Pyramidal indenter* |  | 338 |  |  |
|  | 25% | 113 | 8.99 ± 1.57 (8.50) | 0.36 ± 0.02 (0.10) |
|  | 45% | 85 | 5.85 ± 1.85 (8.71) | 0.12 ± 0.01 (0.03) |
|  | 60% | 98 | 1.75 ± 0.09 (0.44) | 0.19 ± 0.01 (0.04) |
|  | 75% | 42 | 6.99 ± 3.48 (11.49) | 0.11 ± 0.02 (0.06) |
| *Hemispherical indenter* |  | 407 |  |  |
|  | 25% | 91 | 7.34 ± 2.11 (10.27) | 0.32 ± 0.02 (0.10) |
|  | 45% | 74 | 7.70 ± 1.48 (6.51) | 0.21 ± 0.01 (0.06) |
|  | 60% | 157 | 4.57 ± 1.07 (6.83) | 0.17 ± 0.01 (0.03) |
|  | 75% | 85 | 7.89 ± 2.91 (13.68) | 0.14 ± 0.01 (0.05) |

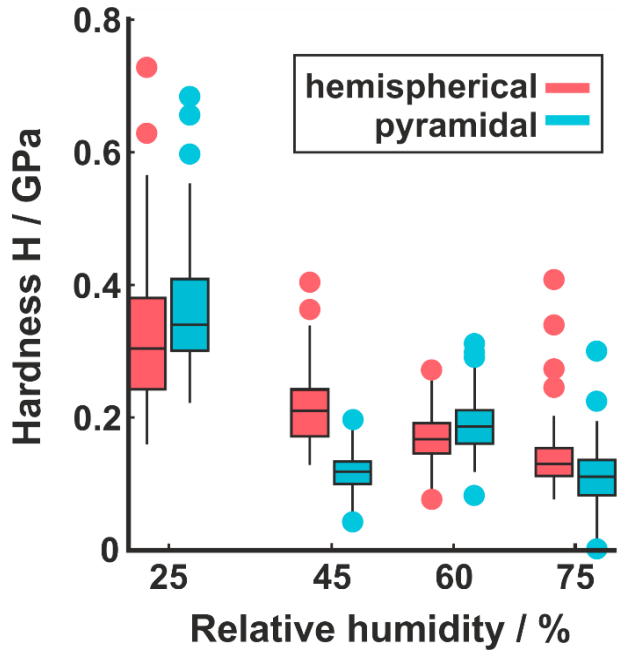

*Figure 7: Results for the hardness of the fibers in longitudinal direction investigated at 25 %, 45 %, 60 %, and 75% relative humidity with both indenter shapes. The horizontal bar indicates the median of the group, the box is the interquartile range, the whiskers cover the range of data not considered outliers and the dots are considered outliers. The criterion to classify an observation as an outlier is that it is greater than $q_3 + 1.5 \cdot (q_3 - q_1)$ or less than $q_1 - 1.5 \cdot (q_3 - q_1)$, where $q_1, q_3$ are the 25$^{th}$ and 75$^{th}$ percentiles of the sample data, respectively.*

The hardness which is presented in Figure 7 also decreases at higher relative humidity levels. The variation close to 50 % relative humidity appears minor, and the largest difference is between 25 % and 45 % relative humidity. The data exhibits only minor skewness compared to the indentation moduli. The results qualitatively agree with the trends in [27].

Since the embedding material is a soft hydrogel-like material, the mechanical properties of GMA were also investigated by AFM-NI using the pyramidal probe. For GMA, the values of $M$ and $H$ are always lower than for the S2 cell wall, and the material's properties also exhibit a dependence on RH. With increasing RH, the values decrease. Due to its hydrogel-like nature, GMA swells a lot and, therefore, measurements in water were not possible. However, the influence of the GMA on $M$ and $H$ of the S2 layer was not significant. The humidity dependence of the indentation modulus and hardness of GMA can be found in the ESI (Figure A2).

In literature, conventional NI has been applied to the S2 wall of spruce wood tracheids [18–21] as well as pulp fibers [24,25]. These results are summarized in Table A1 in the ESI. As in this work, the samples were embedded in resin and cut by a microtome, but all these measurements were obtained under ambient conditions and with varying experimental protocols and algorithm complexity. Comparison of the NI literature data to the results at 45 % relative humidity in this work reveals that the indentation modulus obtained by NI is higher than for AFM-NI. The hardness values of NI measurements range between 250 and 550 MPa [67], and the $H$ values obtained by AFM-NI are similar to the lower limit of this range.

For this reason, additional NI experiments were obtained on the S2 layer of this sample at about 45 % relative humidity for comparison with the AFM-NI results. These experiments were performed using a different sample, different equipment, different operator and different postprocessing algorithms, to minimize the possibility of an error influencing both sets of measurements (the AFM-NI and the NI, respectively). Three different fibers were investigated by NI with altogether 52 indents, resulting in a value for the indentation modulus of 5.75 ± 0.52 GPa (mean ± 1 standard deviation). This is in good agreement with the AFM-NI measurements.

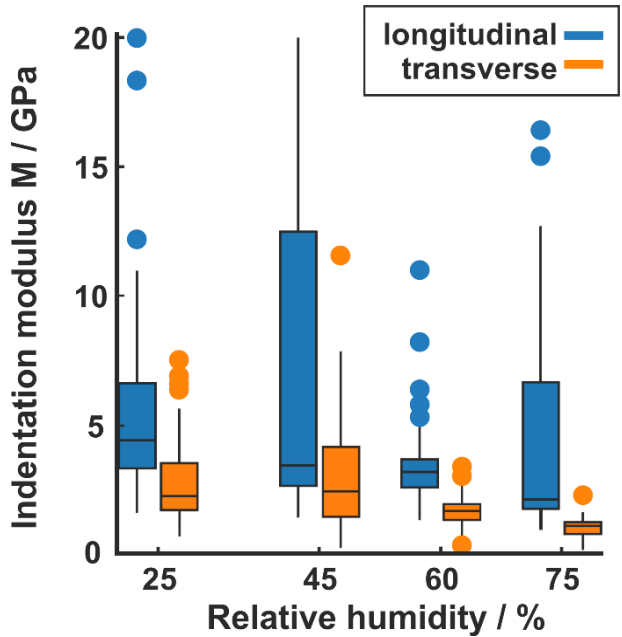

*Figure 8: Results of the indentation modulus in the longitudinal and transverse fiber direction obtained with the hemispherical probe at different relative humidity levels. The horizontal bar indicates the median of the group, the box is the interquartile range, the whiskers cover the range of data not considered outliers, and the dots are outliers. The criterion to classify an observation as an outlier is that it is greater than $q_3 + 1.5 \cdot (q_3 - q_1)$ or less than $q_1 - 1.5 \cdot (q_3 - q_1)$, where $q_1, q_3$ are the 25$^{th}$ and 75$^{th}$ percentiles of the sample data, respectively.*

The low indentation moduli raise questions about the validity of a small strain contact formulation, such as the one used here, since the indentations are performed under force control. Violating the small strain assumption results in an underestimated contact area and overestimated pressure magnitude. At indentation depths $z > qR$ these quantities begin exhibiting unacceptable errors, with different authors putting $q$ around 0.1 [68]. However, the force resultant of the indentation is comparatively insensitive to these errors, as it is calculated by integrating the pressure over the area, resulting in the errors canceling to some extent. Comparing with FEM simulations, Wu et al. found that if the force and the indentation depth are the only quantities of interest, the small strain assumption is acceptable up to $q = 2/3$ for spherical indentation [69]. This conclusion is contingent on the material response being elastic and the indentation field being radially self-similar under the indentation of a unit point load. These assumptions are the same as those made in this work. Our indentation depths at 45 % relative humidity in the longitudinal direction were 59 ± 20 nm (mean ± 1 standard deviation) and 123 ± 23 nm (mean ± 1 standard deviation) with the pyramidal indenter. In the transverse direction, the corresponding data for the hemispherical indenter was 79 ± 43 nm (mean ± 1 standard deviation). This suggests that the indentations were performed below the upper limit for force resultants ($q_L = 59/300 = 0.2, q_T = 79/300 = 0.26$). In the ESI, section S6, the effect of deeper penetration due to the creep during the test is investigated using an FEM model, and the effect is shown to be limited.

### Determining the elastic constants $E_L$ and $E_T$

In previous work [29], the viscoelastic properties of wood pulp fibers in the transverse direction were characterized. For those fibers, AFM-NI measurements in the elastic regime were also obtained at 25 %, 45 %, 60 %, and 75 % relative humidity with a hemispherical probe. It should be noted that for AFM measurements of individual fibers in the transverse direction, mostly the S1 layer is accessible. Therefore, in the results presented in Figure 8, the values for the longitudinal direction are obtained from measurements on the S2 layer and the values for the transverse direction have been measured on the S1 layer. This has some important implications: First, in the S1 layer, the cellulose fibrils are not oriented in the same direction as in the S2. Second, the lignin, hemicellulose, and cellulose content may not be the same in the S1 and the S2. Although the fibrils are not oriented along the axis in the S1, it is generally accepted that they are oriented in the plane of the fiber wall [12]. Hence, an indentation in the transverse direction is still an indentation along an axis normal to the axis of transverse symmetry. Furthermore, the S1 layer is relatively thin (about 200 nm [70]) and is made thinner by chemical treatment. The S1 layer may be sufficiently thin for the indentation measurement to capture mainly the S2 layer response. Such behavior is known when testing thin films [71,72].

The indentations were not performed on the same fibers, and therefore the most appropriate analysis is to use the mean values of the sample groups at 45 % relative humidity for both indenter shapes as inputs when determining the longitudinal and transverse modulus. The value of the $\nu_{TT}$ for the same kind of pulp fibers in this study was determined in [73] using inverse modeling of bulge tests. The value of the shear modulus $G_{TL}$ was determined in [74] using micromechanical single fiber testing. These values have been obtained at a similar relative humidity level and are comparable to the results obtained at 45 % relative humidity in this study.

Table 2: Inputs for the determination of the unknown parameters $E_L$ and $E_T$ based on measurements obtained from the same type of pulp.

| Material property | | Conditions | Source |
|---|---|---|---|
| $M_L$ | Hemi-sphere: 7.70 GPa<br>Pyramid: 5.85 GPa | 45 % RH, 25 °C | This study, mean |
| $M_T$ | Hemi-sphere: 2.92 GPa | 45 % RH, 25 °C | This study, mean |
| $\nu_{TL}$ | 0.25 | | Assumed |
| $\nu_{TT}$ | 0.25 | 50 % RH, 23 °C | [73] |
| $G_{TL}$ | 2.51 GPa | 50 % RH, 23 °C | [74] |
| MFA $\theta$ | 0 | | Assumed |

Without exact knowledge of the MFA, this approach has the advantage of offering a comparatively cheap estimate of the elastic constants. In Figure 9 the relative error contours are presented, indicating that the optimum (determined by Equation (12)) is unique. There are three equations (Equation (8) in the transverse and Equation (7) twice with different indenters) and two unknown parameters ($E_L, E_T$).

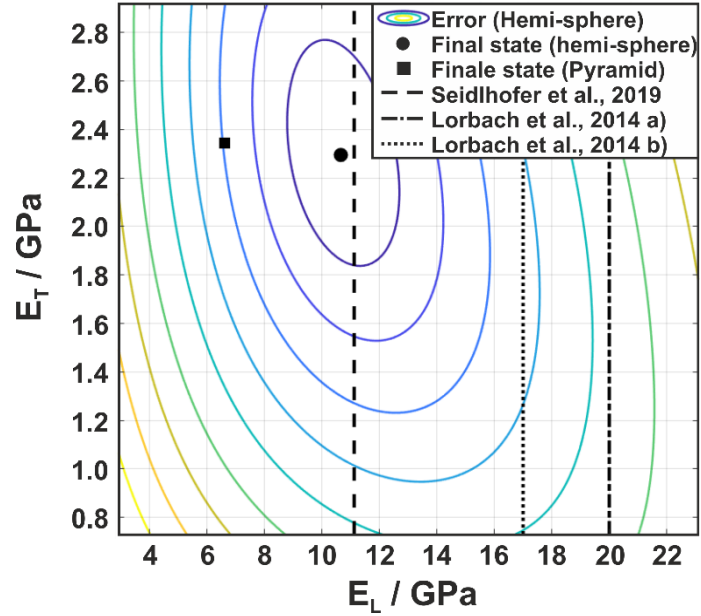

*Figure 9: Error contour and error-minimizing elastic parameters given the indentation moduli measured at 45% relative humidity and assuming the material axis of symmetry is aligned with the normal of the indentation. In various dashed black lines are the mean values of the longitudinal elastic modulus estimated by different micro-mechanical axial single fiber testing procedures in tension on the same type of pulp [75,76].*

The best-fitting parameter combination for the hemispherical indenter is a longitudinal elastic modulus $E_L = 10.7$ GPa and a transverse elastic modulus $E_T = 2.3$ GPa. These values are low compared to previously reported micro- and nano-scale measurements, especially when comparing against the wood. The values are consistent with the magnitude of the shear modulus $G_{TL} = 2.5$ GPa derived using micro-mechanical testing (also about half of what Jäger et al. found for the fibers examined in [51]) and about the same as the longitudinal elastic modulus obtained using single fiber tensile testing for the same kind of pulp fibers in a recent study $E_L = 11$ GPa). For the pyramidal indenter, a similarly unique optimum was obtained, but with a value of $E_L = 6.7$ GPa which is far lower than any reported value in literature. Another study of the same type of pulp fibers – also utilizing single fiber tensile testing and dynamic mechanical analysis – obtained results with mean values of $E_L = 20$ GPa and $E_L = 17$ GPa, respectively [76].

### *Comparing the obtained longitudinal modulus with the sheet modulus*

Six in-plane isotropic handsheets were produced, cut into four test specimens per sheet, and tested in tension. The nominal density of the sheets (grammage divided by average thickness) and elastic modulus are reported in Table 3.

*Table 3: Mechanical properties of tested handsheets and fiber density relevant to Cox' equation (Equation (13)). Measured properties given as mean ± 1 standard deviation.*

| | | | |
|---|---|---|---|
| $E^{(sheet)}$ | 1.55 ± 0.07 | GPa | Measured, n = 24 |
| $\rho^{(sheet)}$ | 459 ± 10.0 | kg m$^{-3}$ | Measured, n = 6 |
| $\rho^{(fiber)}$ | 1250-1500 | kg m$^{-3}$ | Assumed based on p. 788 in [12] |

Solving Equation (13) for the longitudinal elastic modulus of the fiber yields

$$E_L^{(fiber, Cox)} = 3E^{(sheet)} \frac{\rho^{(fiber)}}{\rho^{(sheet)}} = 15.2 \text{ GPa} \qquad (14)$$

As Cox theory is an upper limit for the sheet modulus, when inverted it becomes a lower limit for the fiber modulus. This prediction is 30 % higher than the prediction from AFM-NI indentations – using the hemi-spherical indenter – and closer to the results of Lorbach et al. [76]. The only significant uncertainty is the density of the fiber. Here, the isolated wood fiber density was used as is customary. However, the removal of lignin during the pulp preparation phase certainly decreases this value and increases the pore volume in the fiber. Taking 1250 kg m$^{-3}$ as a lower bound for the fiber density after delignification, the lower bound of the longitudinal fiber modulus is predicted to be 12.6 GPa.

## Discussion

The results show that the indentation modulus obtained with AFM-NI experiments on wood pulp fibers inside the sheet is lower than has been previously reported for single isolated fibers and tested either by nano-indentation or by micromechanical tensile tests. However, the results are in line with the values found from previous studies of fibers extracted from freely dried handsheets [7,17].

The indentation modulus decreases with increasing relative humidity, which is in agreement with previous results [32]. As commonly observed on biological samples, the scatter in the obtained data is relatively large and necessitated many indentations, especially in the longitudinal fiber direction. Part of this scatter likely originates from the MFA. Since the MFA is unknown, it introduces an offset between the indentation direction and the axis of material symmetry. As the MFA does not play a role in the transverse direction, this may also explain the comparatively low scatter of the indentation moduli obtained for this direction.

The indentation moduli obtained with AFM-NI and NI in this article are significantly lower than previously reported indentation moduli for wood and pulp fibers. There are several possible reasons for this difference:

- ***Low control over the indentation direction and the orientation of the surface normal relative to the axis of symmetry in the S2 layer***. If the MFA was significantly higher than in previous studies, or if the indentation normal was not oriented along the fiber axis, this introduces an error in the calculation of the elastic constants. In effect, the method of Delafargue and Ulm [56] is not applicable for this case. Instead, the analysis should use the uncondensed theory presented by Vlassak et al., where the MFA can be easily incorporated. In the ESI (Figure A3), the sensitivity of the elastic moduli with respect to the MFA is investigated. If the MFA is relatively small (< 20 degrees), the effect should remain small.
- ***Drying conditions***. In this work, the fibers were dried inside a sheet. The NI studies surveyed either address wood fibers or do not discuss the sample preparation. Hence, it is not possible to comment on exact differences in the experimental protocol. However, it is well known that drying under restraint has a large influence on the longitudinal elastic modulus of fibers, and that fibers extracted from paper sheets tend to have a region of slack before they resist further elongation [7,17]. However, calculating the longitudinal fiber modulus by inverting the Cox formula, which should yield the lower estimate, also resulted in a higher modulus than that obtained using the indentation. Here, drying conditions are fully accounted for.
- ***Delignification***. In the kraft pulp used for this study most of the lignin was chemically removed. However, the lignin content of the pulp fibers investigated is still higher than that of the pulp fibers measured in [25]. In the absence of buckling phenomena, the increased porosity due to delignification should affect tension and compression tests similarly for small indentations. At deeper indentation, porosity causes a stiffening response compared to in tension when pores close under the indenter. Delignification removes some of the mass from the fiber, increasing internal pores. Thus, the density of the fiber decreases. This decrease is not accounted for in Equation (14).

The relatively low values for the indentation moduli resulted in low elastic stiffness constants. This suggests there could be – at least on the nanoscale at which these experiments were performed – some differences in tensile and compressive responses. However, the conventional NI testing resulted in higher elastic constants, close to those obtained in tensile testing of individual fibers. Hence, if some asymmetry in response does exist, it is unclear if it matters on the scale of the fiber.

# Conclusions & Outlook

AFM-NI was used to determine the indentation modulus along the longitudinal and the transverse axes of wood pulp fibers at different relative humidity levels. These measurements were performed with two different indenter shapes. Since the samples were prepared from microtome-cuts of paper sheets, the indentation moduli can be considered as the in-situ moduli of the fibers inside the paper sheet. The indentation modulus was low compared to values that have been obtained from previous NI experiments for wood and wood pulp fibers in the literature, whereas the hardness was comparable. However, a comparison of the AFM-NI results with NI measurements on the same sample resulted in similar indentation modulus values.

Since wood pulp fibers are anisotropic, the measured indentation modulus is a mix of stiffness tensor components. Therefore, the obtained indentation moduli were used together with inverse modeling to estimate the longitudinal and transverse elastic modulus of the fibers. These values were also lower than those reported in the literature using other methods.

In future work, the determination of the MFA will be an important step to reduce uncertainties and increase the reliability of the results. Furthermore, a non-contact optical method such as Brillouin light scattering spectroscopy could bring more insight into the mechanical behavior of wood pulp fibers [77]. With this technique, it is possible to not only access the longitudinal and transverse direction but also the shear response [78,79] which has not been accessible for wood pulp fibers with a single method before.

The work presented has demonstrated the variability in outcome using multiple indenters to obtain two estimates of the indentation moduli in the longitudinal direction. Although limited uncertainty analyses exist in the literature [42,46], we are unaware of an uncertainty analysis spanning the range from indentation to stiffness characterization. Such a framework would be useful to differentiate exploratory analysis from hypothesis testing.

In the future, the presented method could be used to track the effect of macroscopic constraints on the development of elastic properties of the fibers depending on orientation. Previous work using micromechanical testing [17] indicates a large dependence on the orientation of the fiber relative to the direction of the constraint but did not reach statistical significance due to the low number of samples tested. Industrial experience confirms this dependency, and fiber network modeling [80][81][82] suggests it is comparable to the effect of fiber orientation anisotropy. A customized investigation could confirm these results and attempt to confirm the exact functional form.

## Acknowledgments

The financial support by the Austrian Federal Ministry for Digital and Economic Affairs and the National Foundation for Research Technology and Development is gratefully acknowledged. We also thank our industrial partners Mondi Group, Canon Production Printing, Kelheim Fibres GmbH, SIG Combibloc Group AG for fruitful discussions, and their financial support. The Raman-TERS lab at Montanuniversitaet Leoben is acknowledged for infrastructural support. Special thanks to Angela Wolfbauer of the Institute of Bioproducts and Paper Technology for sample preparation.

## Data Availability

The raw (unprocessed) data used in this work is available in the Zenodo repository. The code necessary to produce the data analysis (Figure 6 - Figure 9) and the figures in the ESI is available in the Zenodo repository [83][84].

## Conflict of Interests

The authors declare that they have no conflict of interest.

**Electronic Supplementary Information (ESI) for Estimation of the in-situ elastic constants of wood pulp fibers in freely dried paper via AFM experiments**

**S1: Experimental AFM-NI curves**

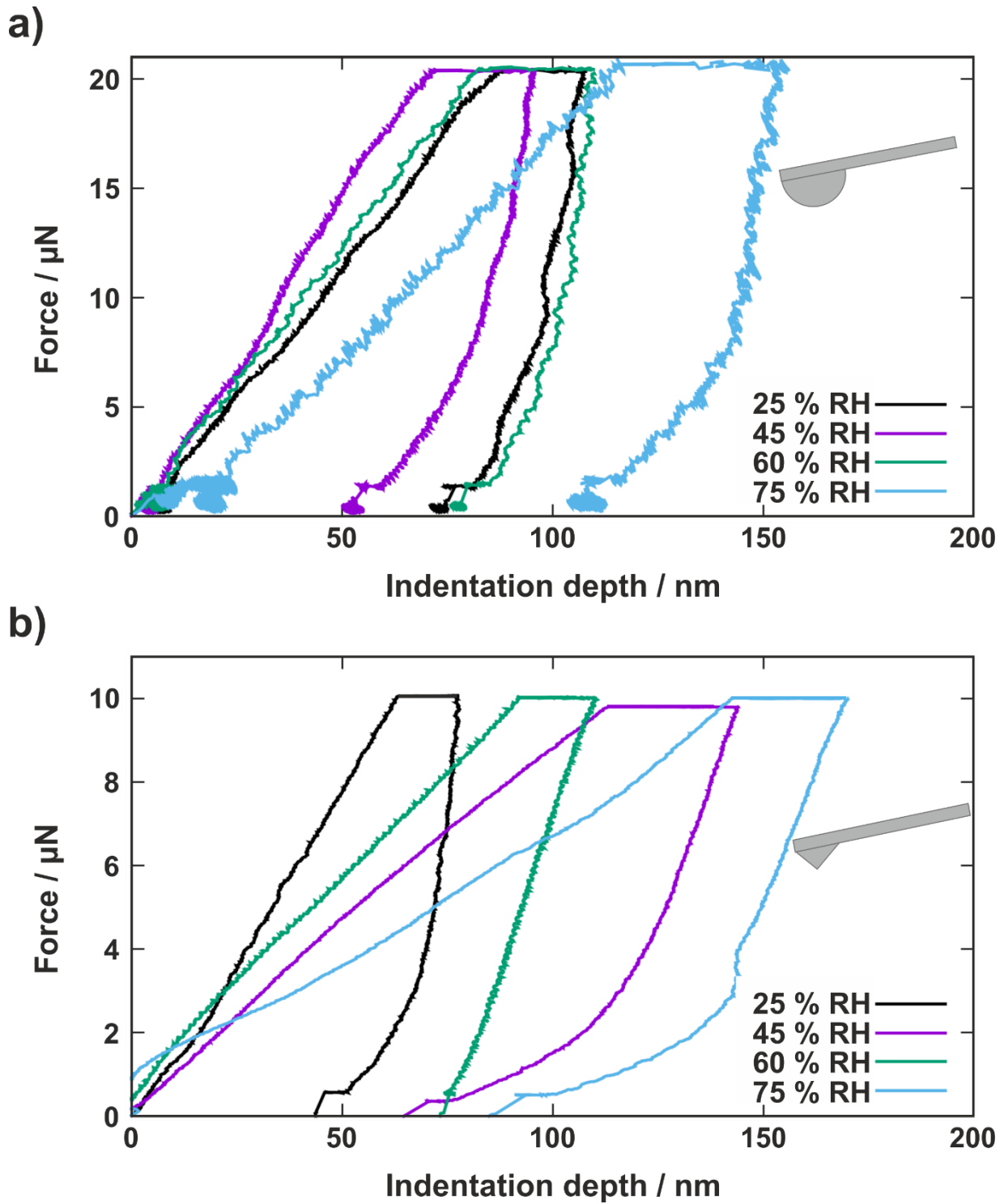

*Figure A 1: Representative force-indentation depth plots of the AFM-NI experiments for all relative humidity levels for (a) the hemispherical probe, and (b) the pyramidal probe.*

## S2: Mechanical properties of the embedding material glycol methacrylate (GMA)

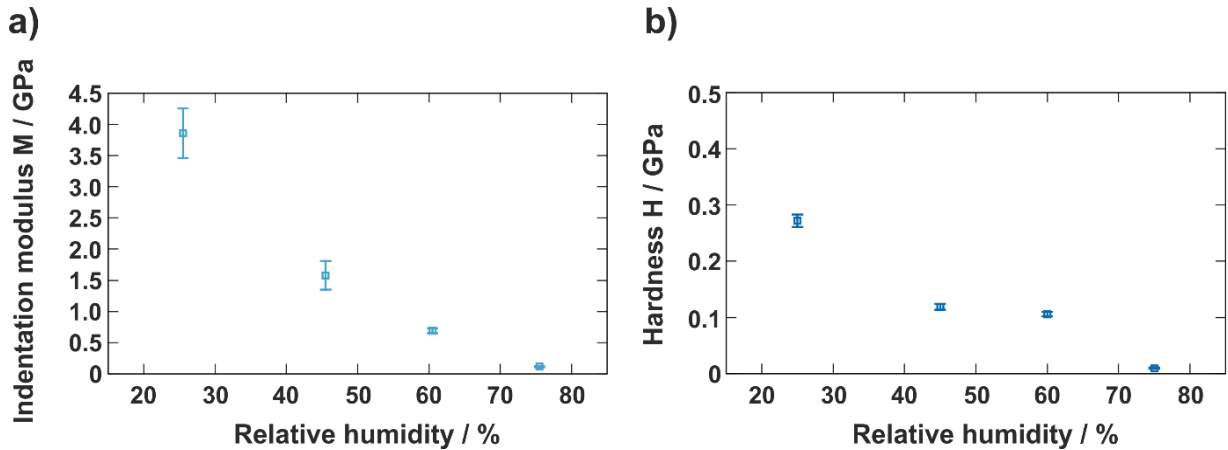

*Figure A 2: (a) Indentation modulus M and (b) hardness H of glycol methacrylate (GMA) obtained with the pyramidal probe at different relative humidity levels.*

## S3: Overview of literature results for the S2 layer

*Table A 1: Overview of the literature values of conventional nanoindentation (NI) measured on the S2 layer of wood and pulp fibers. Mean values obtained by AFM-NI at 45 % RH with two different probe geometries in this work are highlighted in blue for comparison.*

| wood species | detailed information | MFA | pulping grade | probe | $E_r$ / GPa | H / MPa | Ref. |
|---|---|---|---|---|---|---|---|
| spruce | earlywood (EW) | x | x | Berkovich | 13.49 ± 5.75 | 254 ± 69 | [1] |
|  | transition wood (TW) | x | x |  | 21.27 ± 3.12 | 286 ± 39 |  |
|  | latewood (LW) | x | x |  | 21.00 ± 3.34 | 335 ± 30 |  |
| spruce | compression wood (CW) | 50 | x | Berkovich | 8.2 | ~450 | [2] |
|  | EW/LW | 0 | x |  | 17.1 | ~450 |  |
| spruce | LW | 3 | x | Berkovich | 15.81 ± 1.61 | x | [3] |
|  | LW | 5 | x |  | 15.34 ± 0.4 |  |  |
|  | LW | 5 | x |  | 17.08 ± 0.55 |  |  |
|  | TW | 7.5 | x |  | 13.46 ± 0.30 |  |  |
|  | TW | 7 | x |  | 17.54 ± 0.37 |  |  |
| eucalyptus | pulp | x | Unbleached (κ=13) | cube corner | 12.2 ± 1.6 | 420 ± 50 | [4] |
|  |  | x | Bleached (κ=1.5) |  | 10.5 ± 2.1 | 377 ± 33 |  |
| pine |  | x | Unbleached (κ=29) |  | 9.1 ± 1.6 | 430 ± 50 |  |
|  |  | x | Bleached (κ<1) |  | 10.5 ± 2.1 | 377 ± 33 |  |
| spruce/ pine | pulp | x | unbleached (κ<10) | Berkovich | ~16 | ~420 | [5] |
|  |  | x | bleached (TCF) |  | ~12 | ~250-400 |  |
|  |  | x | bleached + refined |  | ~9 | ~200 |  |
| spruce | pulp | x | unbleached, unrefined kraft pulp (κ = 42) | pyramidal | ~ 5.9 (45 % RH) | ~ 120 (45 % RH) | [This work] |
|  |  |  |  | hemi-spherical | ~ 7.7 (45 % RH) | ~ 210 (45 % RH) |  |

## S4: Influence of the microfibrillar angle (MFA)

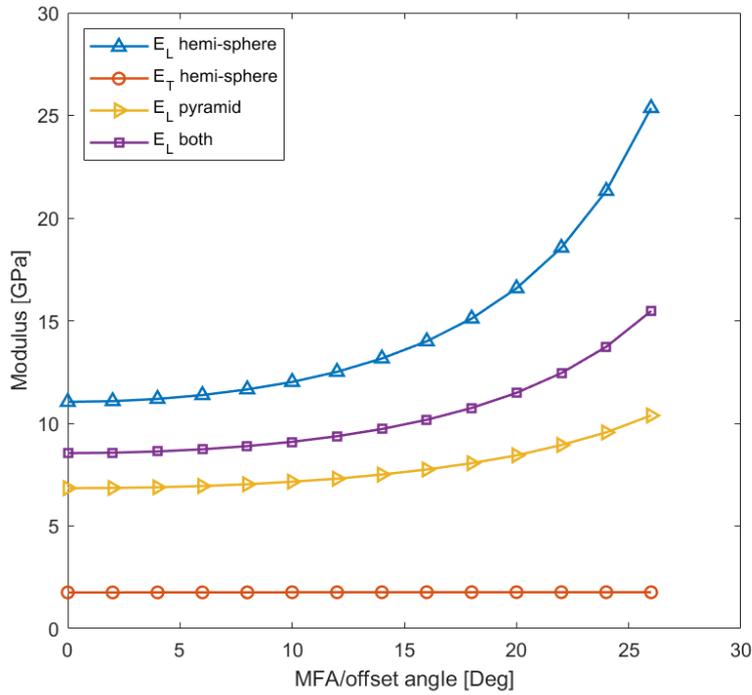

*Figure A 3: Estimate of the elastic constants under different assumptions regarding the orientation of the axis of symmetry. The transverse direction is always assumed to be orthogonal to the microfibrils, as the fibrils are known to never point radially outwards from the lumen of the fiber. Three estimates are presented: The longitudinal modulus fitted using only the hemi-spherical mean value, the pyramid mean value, and both (equally weighted). The range of orientations, 0 – 35 degrees from the longitudinal indentation direction, is realistic for wood pulp fibers.*

## S5: Comparison of different indenters

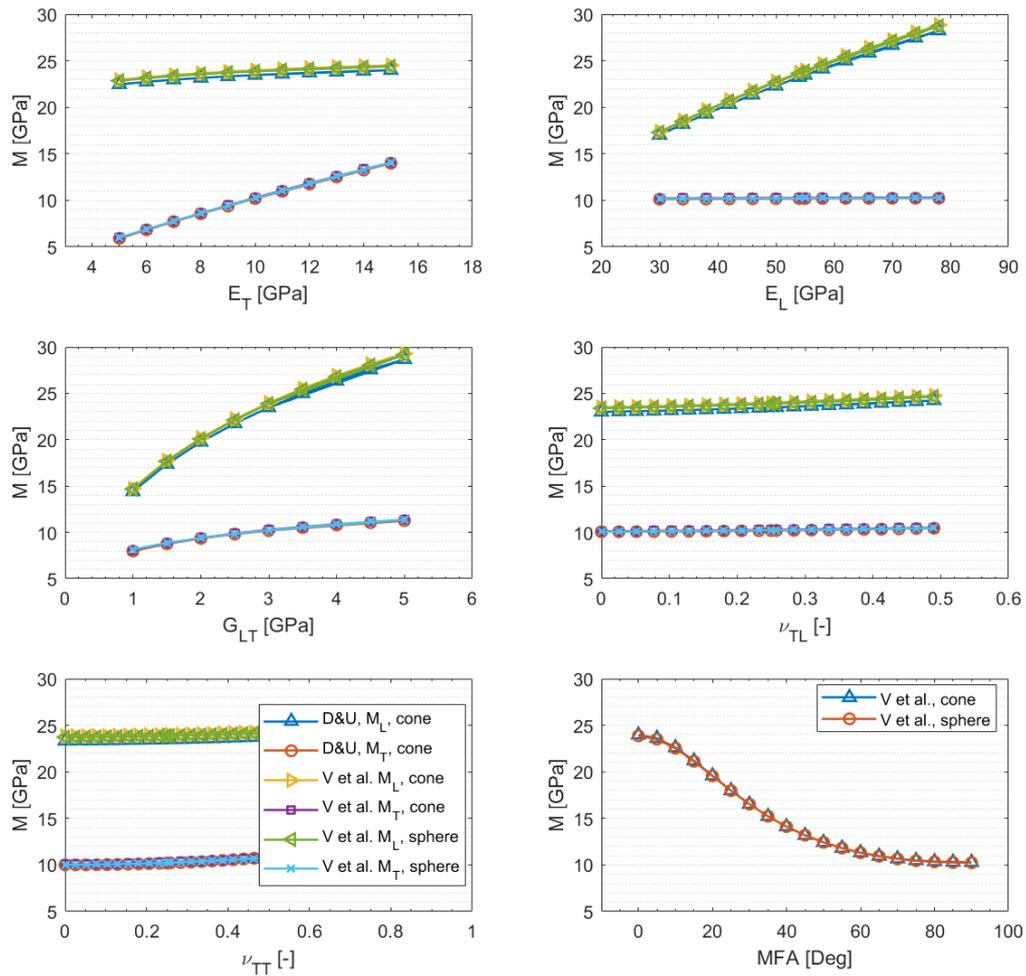

*Figure A 4: Variation in output depending on the method of determining the indentation modulus for the direction parallel with and perpendicular to the axis of symmetry. Shown is the theory presented by Vlassak et al. (V et al.) [6] for the case of a cone and a sphere, respectively, as well as the explicit solution of Delafargue and Ulm (D&U) [7]. Although there is a small difference, the results were considered close enough that the Delafargue and Ulm solution, derived for cones, is also valid.*

## S6: Verification using the finite element method

A finite element model was constructed to evaluate the effect of the assumptions of the applied analytical model. The assumptions in the analytical framework tested are:

- i) Small strain theory is applicable and sufficient.
- ii) The contact is frictionless.
- iii) Viscoelasticity does not significantly affect the indentation modulus.

**Method**

The commercial solver ANSYS was used, and a model was parametrically generated with the intention of matching the experimental conditions. The model employs double symmetry, explicitly modeling only 1/4$^{th}$ of the half-space and 1/4$^{th}$ of the indenter. This somewhat limits the orientations of the material basis that are admissible when comparing with [6] as the major and minor axis of the contact ellipse must lie within the planes of symmetry. The geometry is constructed by sweeping an axisymmetric cross-section 90 degrees around an axis of revolution defined by the intersection of the symmetry planes. The indenter and the half-space are both meshed with homogenous 8-node elements (ANSYS SOLID185). The element coordinate system is set parametrically, enabling the same geometry to be used for indentation along any normal relative to the axis of symmetry. The model is shown in Figure A 5.

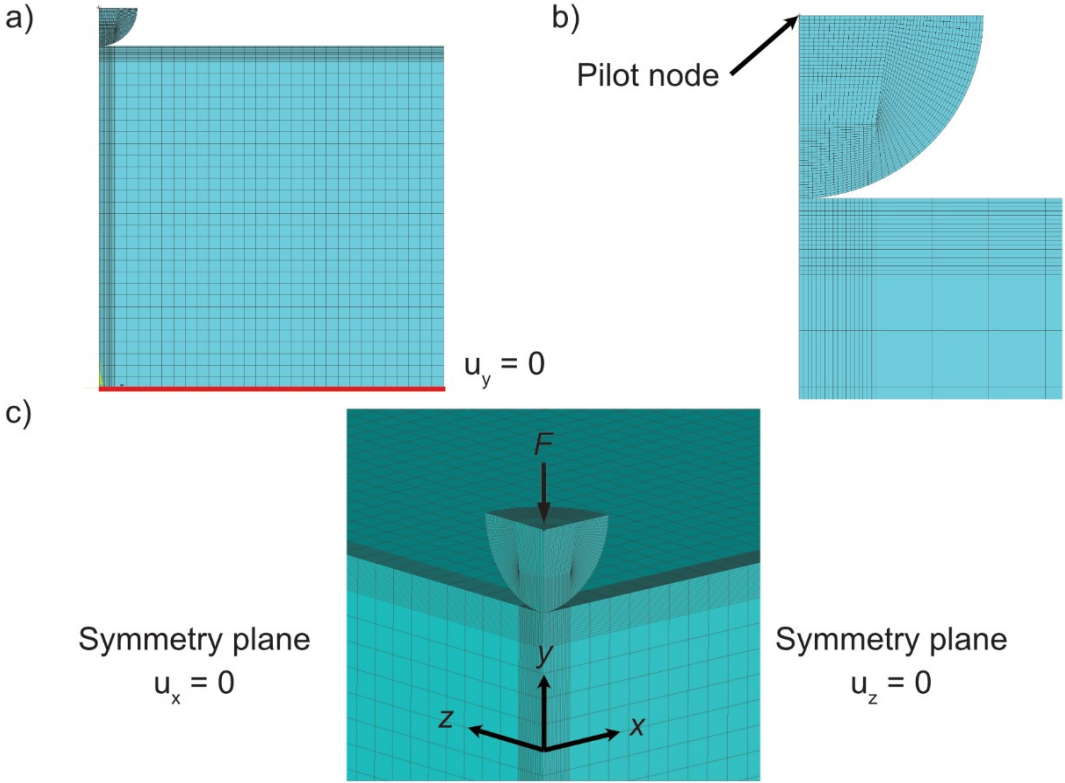

*Figure A 5: FEM model with a) mesh and geometry; b) discretization in the contact zone and indication of the pilot node where the load is applied; c) coordinate system, mesh, and symmetry planes.*

The contact interaction is modeled using a pair-based surface to surface contact (ANSYS CONTA173+TARGE170), with the contact elements on the indenter and the target elements on the half space. The Augmented Lagrange method is used with mortar (projection based) contact detection settings. Contact damping is used in the first time step only. The contact stiffness is set to 10 times the program default.

The necessary symmetry constraints are imposed on the sides of the model. The model is solved using force control and the force is applied via a pilot node located on the top of the indenter acting on the plane of the indenter, such that rotation of the indenter is inhibited.

Large deformation (ANSYS NLGEOM command), the friction coefficient, and the inclusion of viscoelastic behavior is set parametrically.

Any material model implemented in ANSYS can theoretically be examined. Here, we constrained ourselves to transversely isotropic elastic materials with the axis of symmetry set parametrically. In addition, viscoelasticity can be optionally added by supplying a prony series for the bulk and shear response. Note that in ANSYS, general anisotropic viscoelasticity is not implemented for small strain elasticity, only for hyperelastic (ANSYS AHYPER) models.

The viscoelasticity is defined using a prony series for shear only, with a single element in the series, as given in Equation (0.15) where $\alpha_1 = 1.4, \tau_1 = 10\ [s]$ was used, after calibration.

$$G(t) = G \cdot [\alpha_0 + \alpha_1 \exp(-t/\tau_1)] \tag{0.15}$$

**Evaluation**

Verification of the FEM model against the analytical method used in this work is presented in Figure A 6. Throughout, we stuck with the definition of the Poisson's ratios used by A. Jäger et al. in [8].

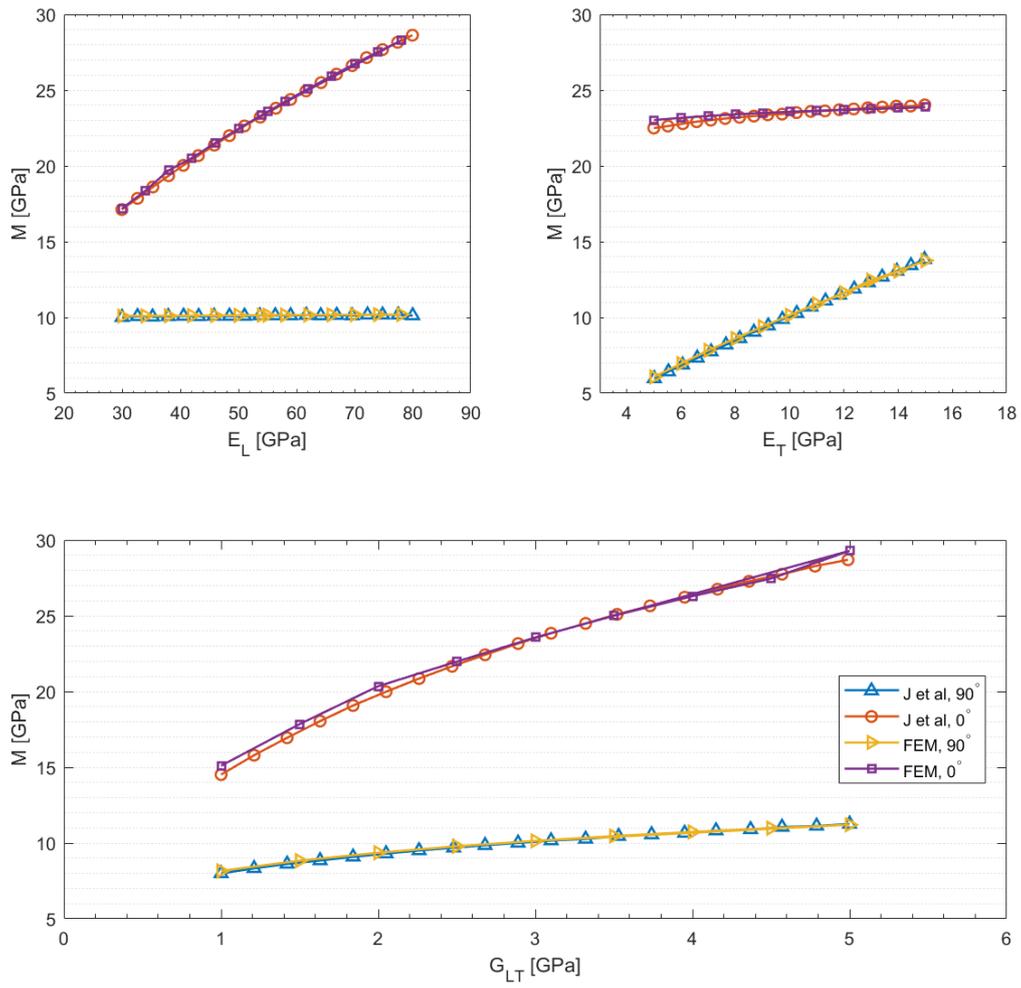

*Figure A 6: Comparison between FEM model and Jäger et al.'s (J et al) results.*

The addition of viscoelasticity resulted in a creep during the hold period, as presented in Figure A 7.

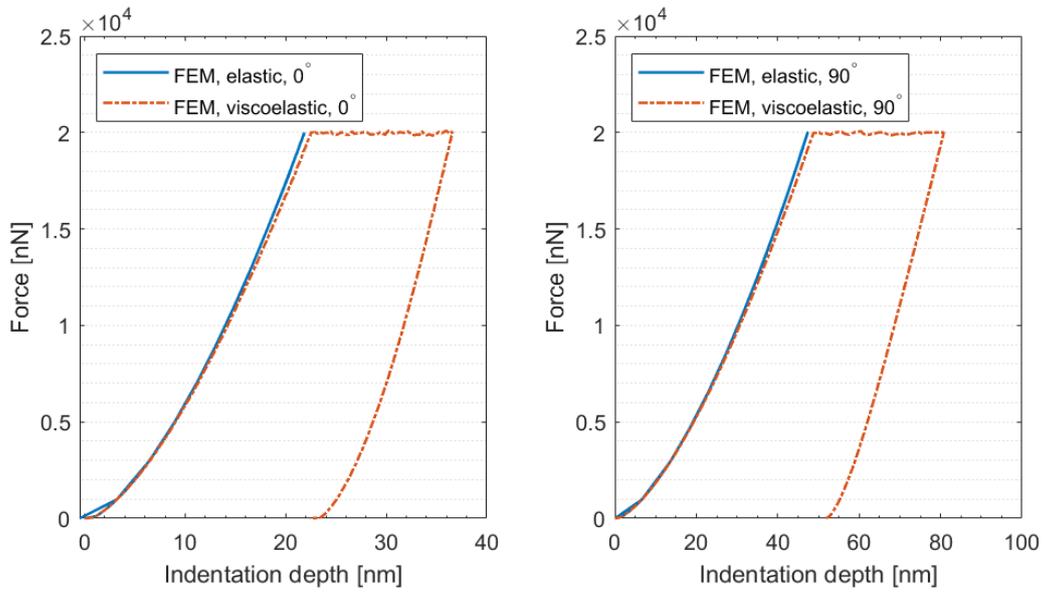

*Figure A 7: Inclusion of a significant amount of viscoelasticity in the model. Left image: Indentation aligned with the axis of transverse material symmetry. Right image: Indentation transverse to the axis of material symmetry. 0 degrees corresponds to the longitudinal direction, and 90 degrees corresponds to the transverse direction.*

After calibration, viscoelasticity, friction, and large deformations can be toggled on and off. The effects of non-linearity and friction are presented in Figure A 8. Here, non-linearity does show an effect, but it is not dominant, whereas friction seems to barely matter.

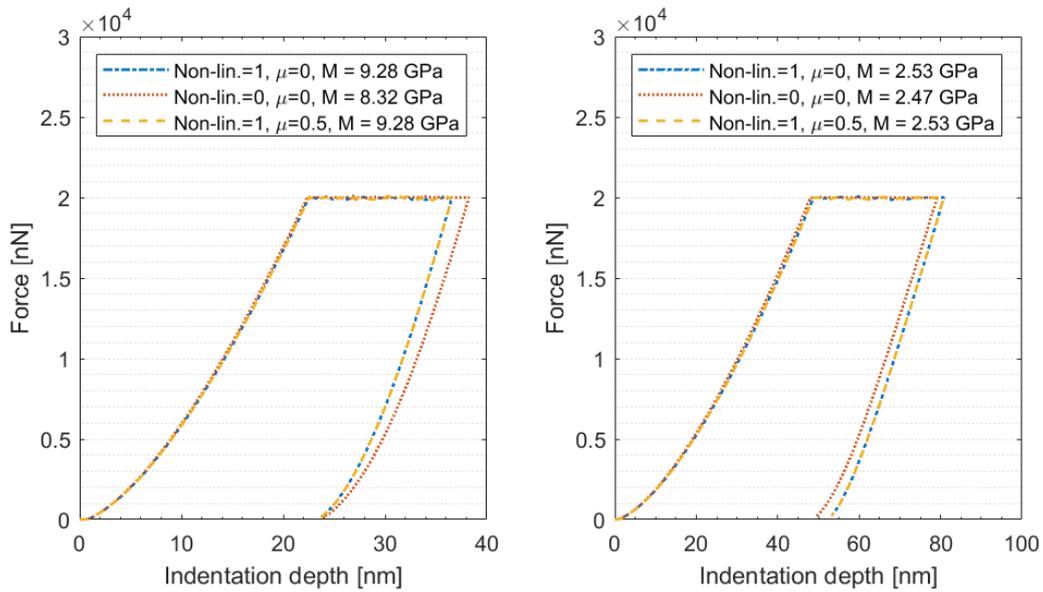

*Figure A 8: Investigation of the effect of nonlinearity and friction in the model. Left image: Indentation aligned with the axis of transverse material symmetry (the longitudinal direction). Right image: Indentation transverse to the axis of material symmetry (the transverse direction).*

**Conclusions**

The investigation using a FEM model shows that:

- The analytical approach is sound.
- The code used to evaluate the indentation modulus based on indentation force-displacement curves is correct (the same code was used to postprocess the FEM model).
- The effect of viscoelasticity is not small but not dominant, at least for the chosen representation of viscoelasticity.
- The effect of non-linearity is not small but not dominant, at least for the investigated setting.
- The effect of friction is small.

Furthermore, it should be noted that the viscoelasticity tended to increase the indentation modulus somewhat, as did the use of a non-linear formulation. A much more rigorous experimental scheme is necessary to draw positive conclusions from the model, but at the minimum it shows that the assumptions made in the manuscript are reasonable.